# LIENARD-WIECHERT POTENTIALS AND METHOD OF IMAGES IN RF FREE ELECTRON LASER PHOTOINJECTOR

W. Salah; The Hashemite University, Zarqa 13115, Jordan, R.M. Jones; Cockcroft Institute, Daresbury, WA4 4AD, UK; University of Manchester, Manchester, M13 9PL, UK.

*Abstract*
Based on Lienard-Weichert retarded potentials and the potential due to the image of charges on the cathode, a rigorous relativistic description of the beam transport inside the RF-photoinjector is presented. The velocity dependent effects are taken into account. Simulations are presented for parameters of the "ELSA" photo-cathode.

# INTRODUCTION

RF-photoinjectors are used as a source of low-emittance and ultra-high brightness electron beams. There are a limited number of codes which take wakefield effects into account in computing electron transport in photoemission. Although there are notable exceptions which do include these effects numerically [1]. The electromagnetic wake in a photoinjector is different from the standard case of a coasting ultrarelativistic beam due to the rapidly changing velocity. In this situation the influence of the acceleration-radiation field, or retardation, must be taken in to account in addition to the image charges on the cathode.

The aim of the present paper is to treat the wakefield of an intense electron beam strongly accelerated inside a cylindrical cavity similar to that of a photoinjector. We employ both Lienard-Weichert potentials and the method of images in order to derive an analytical expression for the field driven by the beam. Electromagnetic field expressions are computed for the "ELSA" photoinjector facility [2] schematized in Fig. 1. Furthermore, by applying the principle of causality we are able to simplify the effects associated with the actual cavity, illustrated in Fig. 1, to an analysis of the electromagnetic fields in a pill-box cavity.

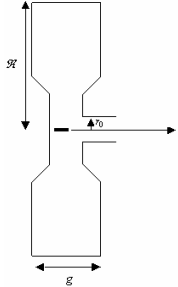

Figure 1: "ELSA" photoinjector (144 MHz cavity).

The beam pulse is assumed to be axisymmetric, of radius $a$, emitted by the cathode from $t=0$ to $t=\tau$ (where $\tau$ is the time at which the photoemission ends), with a constant and uniform current density J. The acceleration RF-electric field $\vec{E}_0$ may be considered as constant and uniform provided, the beam pulse duration

$\tau \ll 1/\nu$ (where $\nu$ is the RF frequency) and the beam radius $a$ is small compared to the cavity radius $\Re$. For the "ELSA" photoinjector $\nu = 144$ MHz, $\Re = 60$ cm, $\pi a^2 = 1$ cm$^2$; the first condition provides the pulse duration $\tau \ll 7$ ns. Under these conditions, the beam velocity $\vec{\beta}(z,t)$ and acceleration $\vec{\eta}(z,t)$ can be shown [3] to be parallel to $\vec{E}_0$ and independent of time:

$$\vec{\beta}(z,t) = \beta(z)\,\vec{u}_z \qquad (1)$$

$$\vec{\eta}(z) = \eta(z)\,\vec{u}_z$$

$$\beta(z) = \frac{\sqrt{(1+Hz(t))^2 - 1}}{1+Hz(t)} \qquad (2)$$

$$\eta(z) = 1+Hz \qquad (3)$$

$$z(t) = \frac{1}{H}\left(\sqrt{1+(Hc(t-t_z))^2} - 1\right) \qquad (4)$$

$$H^{-1} = \frac{mc^2}{eE_0} \qquad (5)$$

where: $m$ and $e$ are the rest mass and charge of the electron, respectively, $z(t)$ is the longitudinal coordinate of an electron at time $t$, and $t_z$ is the time at which an element z of the beam leaves the photocathode.

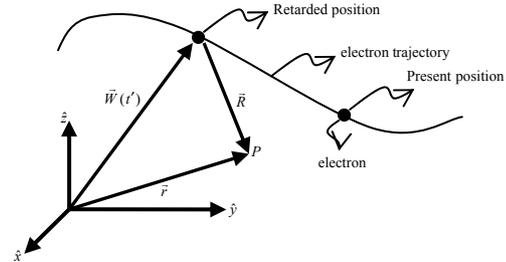

Figure 2: Field driven by an electron.

The electromagnetic fields $(\vec{E},\vec{B})$ generated at time $t$ and point $P$, by an electron, that is moving on a specified trajectory depend on the position W(t') of the electron at time t' (Fig. 2). These fields are driven from the rebuilt scalar and vector potentials $\Phi$ and $\vec{A}$, respectively. Taking into account the boundary condition imposed on the cathode by the equipotential and causality, these fields are given by Lienard-Weichert expression as

$$\vec{E}(P,t\mid W) = -\frac{A}{c(R-\vec{R}\cdot\vec{\beta}(t'))^3}\left(c\frac{(\vec{R}-\vec{\beta}(t')R)}{\gamma(t')} + \vec{R}\times\{(\vec{R}-\vec{\beta}(t')R)\times\frac{\partial\vec{\beta}(t')}{\partial t'}\}\right) \qquad (6)$$

$$\vec{B}(P,t|W) = \frac{1}{c}\frac{\vec{R}}{R} \times \vec{E}(P,t|W) \quad (7)$$

$$\gamma(t') = \left(1 - \beta(t')^2\right)^{-1/2} \quad (8)$$

where $A=(4\pi\varepsilon_0)^{-1}$, $\varepsilon_0$ is the permittivity of free space, $R = W(t')P = c(t-t')$ is the magnitude of the vector from the retarded position $W$ to the field point $P$, and $t'$ is the retarded time.

The first term in the parentheses in equation (6) is the velocity field while the second one is the acceleration or radiation field. The former falls off as $1/R^2$ while the later falls off as $1/R$.

# DEVELOPMENT OF LIENARD - WIECHERT POTENTIALS

The components of the electromagnetic field driven by an electron within the beam and an image of the charge on the cathode can be obtained by the projection of Lienard-Wiechert fields given by equations (6) and (7) on the axes shown in Fig. 3. This projection is applied in the laboratory frame. The point where we observe the field will be taken as the origin of this frame. The cylindrical coordinates ($s$, $\theta$, $z$) of a $W'$ are defined in Fig. 3.

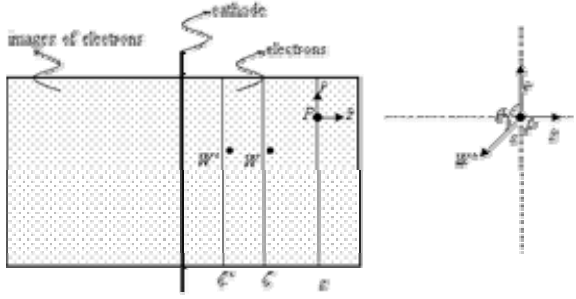

Figure 3: Cylindrical coordinates $s$, $\theta$ and $z$.

The vector from the retarded position of the electron $W(t')$ to the field point $p$ is

$$\vec{R} = \begin{vmatrix} -s\cos\theta \\ -s\sin\theta \\ z - \zeta' \end{vmatrix} \quad (9)$$

where the superscript (′) denotes that the values are taken at time t'. Since a paraxial approximation is used for the beam dynamics, the beam velocity β(t) and the beam acceleration $\frac{\partial \beta(t)}{\partial t}$ are in the same direction. Therefore, the double cross product in equation (6) reads

$$\vec{R} \times ((\vec{R} - \vec{\beta}R) \times \frac{\partial \vec{\beta}}{\partial t}) = -\frac{1}{R^2} \begin{vmatrix} -\frac{\partial \vec{\beta}}{\partial t}s(z-\zeta')\cos\theta \\ -\frac{\partial \vec{\beta}}{\partial t}s(z-\zeta')\sin\theta \\ \frac{\partial \vec{\beta}}{\partial t}s^2 \end{vmatrix} \quad (10)$$

Using equations (10) and (6), the field components $E_z$, $E_r$ and $E_\theta$ on the axes of Fig. 3 are given as

$$E_{z,\beta}(P,t|W) = \frac{A(\zeta'-z+\beta'\sqrt{s^2+(\zeta'-z)^2})}{\gamma^2(\sqrt{s^2+(\zeta'-z)^2}+\beta'(\zeta'-z)^2)^3} \quad (11)$$

$$E_{z,\dot\beta}(P,t|W) = \frac{A\dot\beta's^2}{c(\sqrt{s^2+(\zeta'-z)^2}+\beta'(\zeta'-z)^2)^3} \quad (12)$$

$$E_{r,\beta}(P,t|W) = \frac{As\cos\theta}{\gamma^2(\sqrt{s^2+(\zeta'-z)^2}+\beta'(\zeta'-z)^2)^3} \quad (13)$$

$$E_{r,\dot\beta}(P,t|W) = -\frac{A\dot\beta'(\zeta'-z)s\cos\theta}{c(\sqrt{s^2+(\zeta'-z)^2}+\beta'(\zeta'-z)^2)^3} \quad (14)$$

$$E_{\theta,\beta}(P,t|W) = \frac{As\sin\theta}{\gamma^2(\sqrt{s^2+(\zeta'-z)^2}+\beta'(\zeta'-z)^2)^3} \quad (15)$$

$$E_{\theta,\dot\beta}(P,t|W) = -\frac{A\dot\beta'(\zeta'-z)s\sin\vartheta}{c(\sqrt{s^2+(\zeta'-z)^2}+\beta'(\zeta'-z)^2)^3} \quad (16)$$

where the indices $\beta$ and $\dot\beta = \frac{\partial\beta}{\partial t}$ denote the field components due to the velocity and acceleration; respectively. According to the cylindrical symmetry, the integration of the component $E_\theta$ over the whole beam gives zero.

# GENERATION OF GLOBAL FIELDS FROM INDIVIDUAL COMPONENTS

We generate the global fields driven by the beam using the field components driven by an individual electron and corresponding image charge. For seek of simplicity, we show how we can generate the longitudinal component $E_z$ of the global field, since the other components are identical to the longitudinal one. Consider a cylindrical beam pulse, with radius $a$, carrying a current $I$, emitted by the cathode with a constant and radially uniform current density $J$, moving along the z-axis with velocity β(t) that varies with time. For seek of simplicity, we assume that the shape of the beam does not change during the acceleration. If n($W$, $t$) is the density of electrons or image charge at time $t$ then the longitudinal component of the global field at the point $P$ is

$$E_z(P,t) = \int_D n(W,t) E_z(P,t|W) d^3W + \int_{\bar D} n(W,t) \bar E_z(P,t|\bar W) d^3\bar W \quad (17)$$

where $E_z(P,t|W)$ and $\bar{E}_z(P,t|\bar{W})$ are the field components due to an electron and image charge; respectively, D and $\bar{D}$ represent an ensemble of electrons and image charges; respectively having an antecedent at the retarded time t' and t''.

The components $E_z(P,t|W)$ and $\bar{E}_z(P,t|\bar{W})$ can be written in term of W(t) and $\bar{W}(t)$ using the following

$$W(t') = \Im_{z,t}(M) = \begin{cases} s' = s \\ \theta' = \theta \\ \zeta' = f(s,\theta,\zeta) \end{cases} \quad (18)$$

$$\bar{W}(t'') = \bar{\Im}_{z,t}(\bar{M}) = \begin{cases} s' = s \\ \theta' = \theta \\ \bar{\zeta}' = \bar{f}(s,\theta,\bar{\zeta}) \end{cases} \quad (19)$$

Hence

$$E_z(P,t) = \int_D n(W,t) E_z(P,t|W) \Im_{z,t}(W) d^3W + \int_{\bar{D}} n(W,t) \bar{E}_z(P,t|\bar{W}) \bar{\Im}_{z,t}(\bar{W}) d^3\bar{W} \quad (20)$$

Since the integral will be carried out with respect to $W' = W(t')$ and $\bar{W}' = \bar{W}(t'')$, we can write

$$d^3W = \Omega(\Im^{-1}) d^3\bar{W} \quad (21)$$
$$d^3\bar{W} = \bar{\Omega}(\bar{\Im}^{-1}) d^3\bar{W}' \quad (22)$$

with

$$\Omega(\Im^{-1}) = \frac{\beta}{\beta'}(1 - \frac{\beta'(z-\zeta')}{\sqrt{s^2 + (z-\zeta')^2}}) \quad (23)$$

$$\bar{\Omega}(\bar{\Im}^{-1}) = \frac{\bar{\beta}}{\bar{\beta}'}(1 - \frac{\bar{\beta}'(z-\bar{\zeta}')}{\sqrt{s^2 + (z-\bar{\zeta}')^2}}) \quad (24)$$

where $\Omega(\Im^{-1})$ and $\bar{\Omega}(\bar{\Im}^{-1})$ are the Jacobeans of $\Im^{-1}$ and $\bar{\Im}^{-1}$; respectively.

By means of equations (17-24), equation (17) becomes:

$$E_z(P,t) = A \int_{D(P,\zeta,t)} \frac{J}{e\beta c} (\frac{\zeta'-z+\beta'\sqrt{s^2+(\zeta'-z)^2}}{\gamma'^2(\sqrt{s^2+(\zeta'-z)^2}+\beta'(\zeta'-z))^3} + \frac{\dot{\beta}' s^2}{c(\sqrt{s^2+(\zeta'-z)^2}+\beta'(\zeta'-z))^3})\frac{\beta}{\beta'} \times$$
$$(1 - \frac{\beta'(z-\zeta')}{\sqrt{s^2+(z-\zeta')^2}}) s\, ds\, d\theta\, d\zeta' +$$
$$A \int_{\bar{D}(P,\zeta,t)} \frac{J}{e\bar{\beta}c} (\frac{\bar{\zeta}'-z+\bar{\beta}'\sqrt{s^2+(\bar{\zeta}'-z)^2}}{\gamma'^2(\sqrt{s^2+(\bar{\zeta}'-z)^2}+\bar{\beta}'(\bar{\zeta}'-z))^3} + \frac{\dot{\bar{\beta}}' s^2}{c(\sqrt{s^2+(\bar{\zeta}'-z)^2}+\bar{\beta}'(\bar{\zeta}'-z))^3})\frac{\bar{\beta}}{\bar{\beta}'} \times$$
$$(1 - \frac{\bar{\beta}'(z-\bar{\zeta}')}{\sqrt{s^2+(z-\bar{\zeta}')^2}}) s\, ds\, d\theta\, d\bar{\zeta}' \quad (25)$$

## APPLICATION OF METHOD

We apply the method to several emission regimes [4] from the photocathode. A particularly interesting case is that which occurs at the end of photoemission, corresponding to complete extraction of the beam (i. e at the instant $t = \tau = 30$ ps). This is illustrated in Fig 4. in which the axial electric field is displayed as a function of $Z = Hz$ for the following parameters I =100 A, $E_0$= 30 MV/m. This field is compared to that due to the space charge (or self-field) and the image of charge on the cathode. At the centre of the cathode ($r = 0$, $z = 0$) the beam self field and the field driven by the image of charges on the cathode are similar. However, the field of the beam is dominated by that due to the image charges on the cathode as one moves from the tail to the head of the beam. Far from the cathode the self-field dominates.

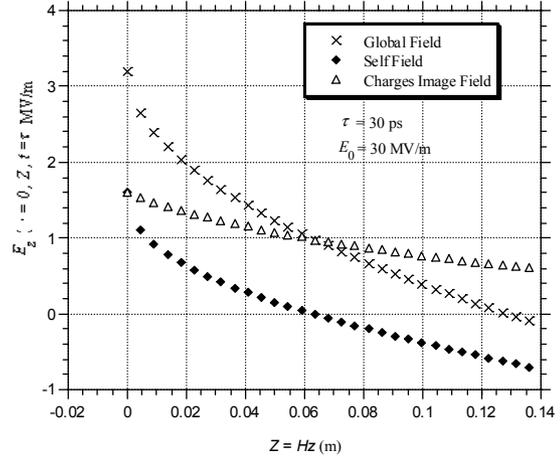

Figure 4: Axial electric field $E_z$ within beam at the end of photoemission.

## ACKNOWLEDGMENT

One of the authors, Wa'el Salah has benefited from a Cockcroft visiting fellowship. The majority of the research presented was completed during the tenure of this fellowship.